\documentstyle[12pt]{article}

\def\be{\begin{equation}}
\def\ee{\end{equation}}
\def\a{\alpha}

\def\G{\Gamma}
\def\d{\partial} 
\def\l{\lambda} 
\def\det{\mbox{det}}
\def\ch{\mbox{ch}}

\def\ln{\mbox{ln}}
\def\exp{\mbox{exp}}
\def\cos{\mbox{cos}}
\def\sin{\mbox{sin}}

\def\ak{a^{+}}
\def\ck{c^{+}}
\def\ra{\rangle}
\def\la{\langle}
\def\D{\Delta}
\def\n{\tilde{n}}
\def\r{\rho}

\begin{document}

\begin{center}
{\bf Asymptotics and functional form of correlators in the 
XX -spin chain of finite length.}
\end{center}
\vspace{0.2in}
\begin{center}
{\large A.A.Ovchinnikov}
\end{center}   
\begin{center}
{\it Institute for Nuclear Research, RAS, Moscow}
\end{center}   
\begin{center}
{\it (May, 2002)}
\end{center}   

\vspace{0.2in}
\begin{abstract}
 We verify the functional form of the asymptotics of the spin-spin 
equal -time correlation function for the XX-chain, predicted by the hypothesis 
of conformal invariance at large distances and by the bosonization procedure.  
We point out that bosonization pocedure also predicts the functional form 
of the correlators for the chains of finite length. We found the exact expression 
for the spin-spin equal-time correlator on finite lattice. 
We find the excellent agreement of the exact correlator with the prediction given 
by the leading asymptotics result up to the very small distances.  
We also establish the correspondence between the value of the constant before the 
asymptotics for the XX-chain with the expression for this constant proposed by 
Lukyanov and Zamolodchikov.  We also evaluate the constant corresponding to the 
subleading term in the asymptotics in a way which is different from the 
previous studies. 
 
\end{abstract}

\vspace{0.5in}

 It is well known that the hypothesis of conformal invariance at large 
distances predicts both the critical exponents and the functional form of 
the correlation functions for massless one - dimensional systems on the circle 
of finite length L (for example see \cite{Cardy}). In this approach the conformal 
mapping $w={L\over2\pi}\ln(z)$ of the infinite plane $z$ on the strip of 
finite width $w=t+ix$, $x\in(0,L)$, is employed. 
It is known that in the case periodic boundary conditions (for the initial spin 
operators) the low-energy theory - is the Conformal Field Theory with cental charge c=1 
(which can be computed for XXZ chain for an arbitrary boundary conditions), 
which is equivalent to the so called Gaussian model of CFT with c=1, with  
known spectrum of primary operators. The only parameter is the compactification radius 
(see below). I would like to stress also that application of CFT for calculation of e.g. 
critical indices is the hypothesis in a sense that it is not proved in a mathematically 
rigorous way. Thus any independent verification of the conformal field theory 
predictions in the exactly solvable models is of interest.  
Alternatively, one 
can use the bosonization procedure for the low-enegry effective theory 
for the systems in the gapless regime (Luttinger liquid) in its different 
versions \cite{ML},\cite{M} to predict the critical exponents for various 
one- dimensional systems (for the first calculation of the critical index for 
the XXZ -spin chain see ref.\cite{LP}). 
In our opinion it is obscure in the literature that 
the bosonization also predicts the functional form of the correlators in the 
form $\sim\cos(2p_Fx)((1/L)\sin(\pi x/L))^{\a}$, where L -is length of the system,  
$\a$ is the critical index and $p_F$ - is the Fermi-momentum, 
at the sufficiently large distances. So it is worth pointing 
out how this functional form appears in the bosonization framework. This is done 
in the first part of the paper. Next, it is interesting to verify this functional 
form in the model where the correlations functions can be calculated exactly. 
In the present paper we study the spin-spin equal-time correlator in the XX - spin 
chain (the hamiltonian is 
$H=\frac{1}{2}\sum_{i=1}^L(\sigma^x_i\sigma^x_{i+1}+\sigma^y_i\sigma^y_{i+1})$, 
the anisotropy parameter $\Delta=0$). 
It is worth mentioning that although the XX-chain can be solved with the help 
of mapping to the free-fermion system via the Jordan-Wigner transformation the system 
is essentially the the hard -core bosons on the lattice (at the half-filling) 
and is not equivalent to the free fermions which manifests for example in the 
completely different correlators in the two models. 
We show in the present letter that the spin-spin correlator in the XX- model can 
be calculated exactly on a chain of finite length at any distance $x$ of order 
of the chain length (see eq.(\ref{sinprod}) below). We show that the functional 
form of the correlator given by (\ref{sinprod}) coincides with that given by the 
conformal fild theory (bosonization).  
Note also that the knowledge 
of the functional form of the correlators can be usefull in the numerical study of 
correlators and extracting the critical exponents in different models. 
In Section 1 we show how the usual bosonization procedure leads to the 
functional form predicted by the conformal invariance. 
In Section 2 we brifly review the well known calculation of the spin -spin 
correlator in the XX -model \cite{LSM}, \cite{McCoy} and present the 
expression for the correlator at the distances $x\sim L$. Finally in Section 3 
we compare the numerical constant in the asymptotic of the correlator given in 
\cite{McCoy} with the expression for this constant proposed recently in 
ref.\cite{Luk} and evaluate the coefficient corresponding to the subleading term in 
the correlator \cite{McCoy}, \cite{Wu} in a new way.

\vspace{0.2in}

{\bf 1. Bosonization.}

\vspace{0.2in}

Consider the effective low -energy Hamiltonian which build up from 
the fermionic operators ($a_k,~c_k,~k=2\pi n/L,~n\in Z$, L- is the length 
of the chain) corresponding to the excitations around the right and the 
left Fermi- points and consists of the kinetic energy term and the 
interaction term $H=T+V$ with the coupling constant $\l$:    
\be
H=\sum_kk(a_k^{+}a_k-c_k^{+}c_k)+\l/L\sum_{k,k',q}\ak_ka_{k+q}\ck_{k'}c_{k'-q}.      
\label{ham}
\ee
Defining the operators \cite{ML}
\[
\r_1(p)=\sum_k\ak_{k+p}a_k,~~~~\r_2(p)=\sum_k\ck_{k+p}c_k,
\]
where $|k|,|k+p|<\Lambda$, where $\Lambda$ is some cut-off energy, 
which for the states with the filled Dirac sea have the following 
commutational relations 
\[
\left[\r_1(-p);\r_1(p')\right]=\frac{pL}{2\pi}\delta_{p,p'}~~~~
\left[\r_2(p);\r_2(-p')\right]=\frac{pL}{2\pi}\delta_{p,p'},
\]
one can represent the Hamiltoniamn in the following form: 
\[
H=\frac{2\pi}{L}\sum_{p>0}\left(\r_1(p)\r_1(-p)+\r_2(-p)\r_2(p)\right)+ 
\l\sum_{p>0}\frac{2\pi}{L}\left(\r_1(p)\r_2(-p)+\r_1(-p)\r_2(p)\right).
\]
First, for the case of free Dirac fermions $\l=0$ the system is equivalent 
to the free scalar Bose field with the Lagrangian 
$L=\frac{1}{2}((\partial_t\phi)^2-(\partial_x\phi)^2)$
\[
 \phi(x,t)= \frac{1}{\sqrt{L}} \sum_k \frac{1}{2\sqrt{|k|}}
\left( b^{+}_k e^{ikx} + b_k e^{-ikx} \right)
\]
where $b_k,b^{+}_k$ are Bose creation and annihilation operators
$\left[b_{k'};b_k^{+}\right]=\delta_{kk'}$ and $kx=|k|t-kx$. 
The free Hamiltonian takes the following form: $H=\sum_p|p|b_p^+b_p$, 
where $p=2\pi n/L$ and the equal time commutation relations are
$\left[\partial_t\phi(x);\phi(y)\right]=-i\delta(x-y)$.
The identification of these operators with the operators $\r_{1,2}(p)$ defined 
above is 
\[
b_p^{+}(p>0)=\r_1(p)\sqrt{2\pi/L|p|}~~~~b_p^{+}(p<0)=\r_2(p)\sqrt{2\pi/L|p|}
\]
for the free fermion case.

To evaluate the correlators in the system of finite length and make the 
connection with the conformal field theory predictions, one can proceed as follows. 
First one can define the lattice fields $n_{1,2}(x)$ with the help of the Fourier 
transform as 
\[
\rho_{1,2}(p)=\sum_x e^{ipx}n_{1,2}(x),~~~~
n_{1,2}(x)=\frac{1}{L}\sum_p e^{-ipx}\rho_{1,2}(p)
\]
This fields have a physical meaning of the local number of the fermions 
above the Fermi level at the right and the left Fermi points. In terms of this 
fields the Hamiltonian has the following form:
\[
H=2\pi\sum_x\left(\frac{1}{2}(n_1^2(x)+n_2^2(x))+\lambda n_1(x)n_2(x)\right)
\]
Considering  the average distribution of the number of extra particles we obtain
the ground state energy in the form
\[
\Delta E=\frac{2\pi}{L}\left(\frac{1}{2}\left((\D N_1)^2+(\D N_2)^2\right)
+\l\D N_1 \D N_2 \right)
\]
where $\D N_{1,2}$ - are the additional numbers of particles at the two Fermi points. 
One can also rewrite the  ground state energy in the sector with the total number of
particles and the momentum $\D N=\D N_1+\D N_2,~~\D Q=\D N_1-\D N_2$
in such a way that the total Hamiltonian takes the following form (this for was first 
proposed in ref.\cite{Haldane}): 
\be
H=u(\l)\sum_p|p|b_p^{+}b_p+\frac{\pi}{2L}u(\l)\left[\xi(\D N)^2+(1/\xi)(\D Q)^2\right],
\label{finite}
\ee
where the parameters $u(\l)=(1-\l^2)^{1/2}$ and $\xi=((1+\l)/(1-\l))^{1/2}$. 
Calculation of finite - size corrections to the energy of the ground state for 
the XXZ- spin chain (see for example \cite{Kar}) leads to the expression  
(\ref{finite}) and allows one to obtain the parameter $\xi$ which leads to 
the predictions of critical indices according to the conformal field theory. 
The calculation gives the value $\xi=2(\pi-\eta)/\pi$, where the parameter $\eta$ 
is connected with the anisotropy parameter of the XXZ - chain as 
$\Delta=\cos(\eta)$. 
Next one establishes the commutational relations for the fields $n_{1,2}(x)$.
We have:
\[
\left[ n_{1}(x); n_{1}(y)\right]=\frac{1}{(2\pi)^2}\frac{2\pi}{L}
\sum_p p e^{ip(x-y)}
\]
To take the continuum limit it is sufficient to extend the sum over the momentum 
to $p\in(\pm\infty)$ (note that initially it was implied the sum in the limits 
$p\in(\pm\pi)$). Then we obtain 
\[
\left[ n_{1}(x); n_{1}(y)\right]\rightarrow 
\frac{1}{(2\pi)^2} \int dp p e^{ip(x-y)}=-\frac{i}{2\pi}\delta'(x-y)
\]
Then intoducing the new variables 
\[
\n_{1,2}(x)=\sqrt{2\pi}~ n_{1,2}(x),
\]
we have the following commutational relations and the density of the 
Hamiltonian 
\be
H=\frac{1}{2}\left(\n_1(x)\n_1(x)+\n_2(x)\n_2(x)\right)
+ \lambda \n_1(x)\n_2(x).
\label{v}
\ee
We also have the following commutational relations
$\left[\n_1(x);\n_1(y)\right]=\delta^{\prime}(x-y)$.   
We then have the following conjugated field and the momenta:
\[
\pi(x)= \frac{1}{\sqrt{2}}(\n_1(x)-\n_2(x));~~~~  
\d_x\phi(x)= \frac{1}{\sqrt{2}}(\n_1(x)+\n_2(x))
\]
\[
\phi(x)= \tilde{N}(x)=\tilde{N}_1(x)+\tilde{N}_2(x),~~~ 
\tilde{N}_{1,2}(x)=\int_0^x dy~\n_{1,2}(y) 
\]
In terms of these variables the Hamiltonian takes the following form: 
\[
H= \frac{1}{2}\int_0^L dx\left((1-\l)\pi^2(x)+ (1+\l)(\d\phi(x))^2\right). 
\]
The Hamiltonian density can also be represented in the following form: 
\be
H=\frac{1}{2}u(\l)\left[ (1/\xi)\pi^2(x)+ \xi(\d\phi(x))^2\right]
 = \frac{1}{2}u(\l)\left[ \hat{\pi}^2(x)+ (\d\hat{\phi}(x))^2\right], 
\label{h}
\ee
where 
\be
\pi(x)=\sqrt{\xi}~\hat{\pi}(x),~~~ 
\phi(x)=(1/\sqrt{\xi})\hat{\phi}(x).  
\label{canon}
\ee
The last equation (\ref{canon}) is nothing else as the canonical transformation,  
which is equivalent to the Bogoliubov transformation for the original operators 
$\rho_{1,2}(p)$. Next to establish the expressions for Fermions one should use 
the commutational relations $\left[a^{+}(x);\rho_{1}(p)\right]=-e^{ipx}a^{+}(x)$
and the same for $c^{+}(x)$. Note that these last relations were obtained using the 
expression with original lattice fermions: 
$\rho_{1}(p)=\sum_y e^{ipy}a^{+}(y)a(y)$. 
In this way we obtain the following expressions for fermionic operators:
\be
a^{+}(x)=K_1~\exp\left(\frac{2\pi}{L}\sum_{p\neq0}\frac{\rho_1(p)}{p}e^{-ipx}\right)
=K_1~\exp\left(-i2\pi N_1(x)\right),
\label{fermions}
\ee
\[
c^{+}(x)=K_2~\exp\left(-\frac{2\pi}{L}\sum_{p\neq0}\frac{\rho_2(p)}{p}e^{-ipx}\right)
=K_2~\exp\left(i2\pi N_2(x)\right),
\]
where the fields $N_{1,2}(x)$ differ by the normalization $\sqrt{2\pi}$ from the 
fields $\tilde{N}_{1,2}(x)$ and $K_{1,2}$ are the Klein factors - the operators 
which creates the single particle at the right (left) Fermi -points  
(we omit here the usual exponential suppression $e^{-\a|p|/2}$ in the exponent and 
and the constant factor in front of the exponent $1/\sqrt{2\pi\a}$ which in the 
limit $\a\rightarrow 0$ leads to the correct anticommutational relations \cite{LP}). 
Note that the above expressions (\ref{fermions}) are equivalent to the known
``field-theoretical'' bosonization formulas \cite{M}: 
\[
a^{+}(c^{+})(x)=K_{1,2}~
e^{-i\sqrt{\pi}\left(\int_{-\infty}^{x}dy\pi(y)\pm\phi(x)\right)}. 
\]
Now let us apply the above formulas to the specific case of the XXZ - spin chain. 
Using the Jordan-Wigner transformation $\sigma^{+}_x=a^{+}_x\exp(i\pi N(x))$,
where $a^{+}_x$ stands for the ``original'' lattice fermionic operator,  
and performing the obvious substitutions $N(x)\rightarrow x/2+N_1(x)+N_2(x)$ and
$a^{+}_x\rightarrow e^{ip_Fx}a^{+}(x)+e^{-ip_Fx}c^{+}(x),~~p_F=\pi/2$,  
we obtain after the canonical transformation (\ref{canon})
the expression for the spin operator which determines 
the leading term in the asympotics of correlator for the XXZ -chain:
\be
\sigma^{+}_x\sim(-1)^x\exp\left(-i\pi\sqrt{\xi}(\hat{N}_1(x)-\hat{N}_2(x))\right),
\label{leading}
\ee
where $\hat{N}_{1,2}(x)$ - are corresponds to the free fields 
$\hat{\pi}(x)$, $\hat{\phi}(x)$, obtained after the transformation (\ref{canon}). 
To these operators correspond the new operators $\rho_{1,2}(p)$ and the new 
fermionic operators (quasiparticles). 
Analogously the term responsible for the subleading asymptotics has the form 
\[
\exp\left(i2\pi(1/\sqrt{\xi})(\hat{N}_1(x)+\hat{N}_2(x))\right)
\exp\left(-i\pi\sqrt{\xi}(\hat{N}_1(x)-\hat{N}_2(x))\right). 
\]
Averaging the product of exponents in bosonic operators 
for the expression (\ref{leading}) and using the properties of $\rho_{1,2}(p)$, 
$\la\rho_1(-p)\rho_1(p)\ra=\frac{pL}{2\pi}\theta(p)$ and 
$\la\rho_2(p)\rho_2(-p)\ra=\frac{pL}{2\pi}\theta(p)$, 
we get for the correlation function 
$G(x)=\la0|\sigma^{+}_{i+x}\sigma^{-}_{i}|0\ra$ the following sum in 
the exponent: 
\[
C~\exp\left(~\frac{\xi}{4}~\sum_{n=1}^{\infty}\frac{1}{n}e^{in(2\pi x/L)}
+ h.c.\right), 
\]
where $C$ - is some constant. Then using the formula   
$\sum_{n=1}^{\infty}\frac{1}{n}z^n=-\ln(1-z)$ and substituting the value
$\xi=2(\pi-\eta)/\pi\rightarrow 1$ we obtain the following 
expression for the XX - chain: 
\be
G(x)=C_0\frac{(-1)^x}{\left(L\sin(\frac{\pi x}{L})\right)^{\alpha}},
~~~\alpha=\frac{\pi-\eta}{\pi}=1/2 ~~~(x>>1).
\label{coas}
\ee
Thus, although bosonization, which deals with the low-energy effective theory, 
is not able to predict the constant before the asymptotics, the critical exponent 
and the functional form are predicted in accordance with conformal field theory.

\vspace{0.3in}

{\bf 2. Asymptotics of correlators in XX -spin chain.}

\vspace{0.2in}

Let us briefly review the exact calculation \cite{LSM} of the spin-spin 
equal-time correlation function (density matrix) for the XX- spin chain on 
finite lattice of the length L: 
\[
G(x)= \la0|\sigma^{+}_{i+x}\sigma^{-}_{i}|0\ra.
\]
Using the Jordan-Wigner transformation relating spin operators to the 
Fermi operators ($a^{+}_i,~a_i$)  
$\sigma^{+}_{x}=\exp(i\pi\sum_{l<x}n_l)a^{+}_{x}$, the correlation function 
$G(x)$ can be represented as the following average for over the free-Fermion 
ground state:
\[
G(x)= \la0|a^{+}_{x}e^{i\pi N(x)}a_{0}|0\ra, 
\]
where $N(x)=\sum_{i=1}^{x-1}n_i$. 
Introducing the operators, anticommuting at different sites,  
\[
A_i=a^{+}_i+a_i, ~~~~~B_i=a^{+}_i-a_i, ~~~~A_iB_i=e^{i\pi n_i}, 
\]
where $n_i=a^{+}_ia_i$- is the fermion occupation number, with the following 
correlators with respect to the free-fermion vacuum, 
\[
\la0|B_iA_j|0\ra=2G_0(i-j),~~~\la0|A_iA_j|0\ra=0,~~~ \la0|B_iB_j|0\ra=0,
\]
where the free-fermion Green function on finite chain $G_0(x)$ is 
\[
G_0(x)=\la0|a^{+}_{i+x}a_{i}|0\ra=\frac{\sin(\pi x/2)}{L\sin(\pi x/L)}, 
\]
one obtains the following expression for the bosonic correlator:
\[
G(x)=\frac{1}{2}\la0|B_0(A_1B_1)(A_2B_2)\ldots(A_{x-1}B_{x-1})A_x|0\ra.
\]
Note that we assume the periodic boundary conditions for the initial spin 
operators, so that strictly speaking, the above formulas are valid only for 
the case when $L$ - is even (the ground state is not degenerate) and 
$M=L/2$ - is an odd integer (so that $L/4$ - is not an integer). In fact it is 
easily seen that for the boundary term $\sim\exp(i\pi(M-1))=+1$ 
when $M$ - is odd, the momentums of fermions - are integer 
(not half-integer), their configuration is symmetric around zero and  
the free - fermionic Green function $G_0(x)$ - is given exactly 
by the above formula. It was also mentioned in ref.\cite{LSM} that the 
periodic boundary conditions for spin operators (``a-cyclic'' in this paper)
leads to the same results as for the periodic boundary conditions for the 
fermionic operators (``c-cyclic'') for the case when $L/4$ - is not an integer.   
Using Wick's theorem we obtain the following determinant of $x\times x$ matrix:
\[
G(x)=\det_{ij}(2G_0(i-j-1)), ~~~~i,j=1,\ldots x.
\]
Due to the form of this matrix ($G_0(l)=0$ for even $l$) this determinant 
can be simplified and the following formulas are obtained: 
\[
G(x)=\frac{1}{2}(R_N)^2,~(x=2N),~~~~G(x)=-\frac{1}{2}R_{N}R_{N+1},~(x=2N+1), 
\]
where we denote by $R_N$ the following determinant of the $N\times N$- matrix: 
\[
R_N=\det_{ij}((-1)^{i-j}G_0(2i-2j-1)), ~~~~i,j=1,\ldots N, 
\]
where $G_0(x)$ is the same Green function of free fermions as above. 
Since $R_N$ - is the Cauchy determinant one can obtain the following 
expression for it on the finite chain: 
\be
R_N= \left(\frac{2}{\pi}\right)^N
\prod_{k=1}^{N-1}\left(\frac{(\sin(\pi(2k)/L))^2}{\sin(\pi(2k+1)/L)
\sin(\pi(2k-1)/L)}\right)^{N-k}. 
\label{sinprod}
\ee
This formula is the main result of the present paper. 
Note that we obtained the exact expression for the correlator on the 
finite lattice. 
As it was mentioned above the expression (\ref{sinprod}) for the correlator was 
obtained for $M$ - odd. As for the case $M$ - even, it is clear on general grounds 
that in this case the correlator $G(x)$ will be modified by the terms of order 
$1/L$ at any distance $x$ (that can be proved rigorously, however the proof will 
not be considered in the present letter). 
From the expression (\ref{sinprod}) it is easy to 
obtain the correlator in the thermodynamic limit ($L\rightarrow\infty$)
which is given by the similar product. We consider the analitical evaluation 
of this product for large $N$ in the next Section. For the finite chain it is easy 
to evaluate (\ref{sinprod}) numerically and compare the result with
the asymptotic (\ref{coas}) at $x>>1$ and $x\sim L$. 
One finds numerically that already at the 
sufficiently small distances $x$, where the contribution of the subleading term
$\sim x^{-5/2}$ becomes negligible, the correction to the asymptotic formula 
(\ref{coas}) behaves like $\sim 1/L$ (at the very large $L\sim 10^5$ it is difficult 
to evaluate the product at $x\sim L$ due to the numerical reasons).  
We find numerically that the exact correlator coincides with 
the correlator given by (\ref{sinprod}) with very high accuracy 
up to the very small distances $x\sim 1$. 
One can also consider the limit $L\rightarrow\infty$ in (\ref{sinprod}) and 
using the exact value of $G(x)$ at $x\sim 1$ estimate the value of the 
subleading term which is predicted to be of order $\sim 1/x^{\a+1/\a}$ for 
the XXZ - chain. The coefficient for this term appears to be very small: 
at $x\sim 1$ (see Section 3, equation (\ref{asympt})) so that 
the correlator for the infinite chain 
can be seen in a very good agreement with the asymptotic formula (\ref{coas}) 
even at the very small distances $x\sim 1$. 
The contribution of the next terms is difficult to obtain numerically 
(presumably due to their smallness). The physical reason for the smallness of the 
subleading term is not clear at the moment.

\vspace{0.2in}

{\bf 3. Numerical constant.}

\vspace{0.2in}

Here we calculate the constant $C_0$ in front of the asymptotics of the 
density matrix in the XX -spin chain and derive the representation 
proposed in ref.\cite{Luk} for this case. Using the expressions in the 
thermodinamic limit we calculate the product: 
\[
R_N= \left(\frac{2}{\pi}\right)^N
\prod_{k=1}^{N-1}\left(\frac{(2k)^2}{(2k+1)(2k-1)}\right)^{N-k}. 
\]
Considering the logarithm of $R_N$ we obtain the following sum:
\[
\ln(R_N)=N\ln(2/\pi)+ 
\sum_{p=1}^{\infty}\frac{1}{p}\left(\frac{1}{2}\right)^{2p}
\sum_{k=1}^{N-1}(N-k)\frac{1}{k^{2p}}
\]
Next, we use the following formula for the finite sum 
$\sum_{k=1}^{n}1/k^s=\zeta(s)-\zeta(s,n+1)$ (where $\zeta(s,n)$ -is 
generalized Riemanian zeta- function) \cite{Brychkov}:
\[
\sum_{k=1}^{n}\frac{1}{k^m}=\frac{(-1)^m}{(m-1)!}\left(\psi^{(m-1)}(1)-
\psi^{(m-1)}(n+1)\right),
\]
where $\psi^{(n)}(z)$ is the $n$-th derivative of $\psi$-function 
$\psi(z)=\G^{'}(z)/\G(z)$, which leads to the following 
expression for $\ln(R_N)$: 
\[
\ln(R_N)= \sum_{p=1}^{\infty}\frac{1}{p}\left(\frac{1}{2}\right)^{2p}  
  \frac{1}{(2p-2)!}\psi^{(2p-2)}(1)
\]
\be
-N \sum_{p=1}^{\infty}\frac{1}{p}\left(\frac{1}{2}\right)^{2p}
 \frac{1}{(2p-1)!}\psi^{(2p-1)}(N)
-\sum_{p=1}^{\infty}\frac{1}{p}\left(\frac{1}{2}\right)^{2p}
\frac{1}{(2p-2)!}\psi^{(2p-2)}(N),
\label{terms}
\ee
since the sum 
\[
N \sum_{p=1}^{\infty}\frac{1}{p} \left(\frac{1}{2}\right)^{2p}  
  \frac{1}{(2p-1)!}\psi^{(2p-1)}(1) = -N\ln(2/\pi)
\]
cancells the term coming from the factor $(2/\pi)^N$. 
The last formula can be obtained using 
$\psi^{(n)}(1)=(-1)^{n+1}n!\zeta(n+1)$ and rewriting the last sum
using the definition of $\zeta$- function as 
$-N\sum_{k=1}^{\infty}\ln(1-1/4k^2)=-N\ln(2/\pi)$, which can be seen from 
the infinite product representation 
$\sin(x)/x=\prod_{k=1}^{\infty}(1-\pi^2x^2/k^2)$. 
Thus we are left with three terms in eq.(\ref{terms}). To find their  
asymptotic behaviour at large $N$ up to the terms of order $\sim 1/N^{2}$, 
we use the asymptotics  
\[
\psi(z)=\ln(z)-\frac{1}{2z}-\sum_{n=1}^{\infty}\frac{B_{2n}}{2nz^{2n}}, 
~~~|z|\rightarrow\infty, 
\]
where $B_{2n}$ - are Bernulli numbers ($B_{2}=1/6$). One can see that the sum 
of the second and the third terms in eq.(\ref{terms}) is 
\be
-\frac{1}{4}\ln N -\frac{1}{4}-\frac{1}{64~N^2} + O\left(\frac{1}{N^4}\right), 
\label{log}
\ee
while the first term equals 
\[
- \frac{1}{4}\gamma + \sum_{p=2}^{\infty}
\frac{1}{p}\left(\frac{1}{2}\right)^{2p}\frac{1}{(2p-2)!}\psi^{(2p-2)}(1), 
\]
where $\gamma=-\psi(1)=0.5772\ldots$ - is the Euler's constant. 
To estimate this sum one can use the integral representation: 
\[
\psi(z)=\int_0^{\infty}dt\left(\frac{e^{-t}}{t}-\frac{e^{-zt}}{1-e^{-t}}
\right).
\]
One also readily obtains the integral representation for $\psi^{(2p-2)}(1)$ 
which leads to the following expression for the first term: 
\[
\frac{1}{4}\int_0^{\infty}dt \left( \frac{e^{-t}}{t}-\frac{e^{-t}}{1-e^{-t}} 
\left( \sum_{p=1}^{\infty}\frac{1}{p}\frac{1}{(2p-2)!}
\left(\frac{t}{2}\right)^{2p-2} \right) \right). 
\]
Thus combining all terms after the simple algebra, using the integration 
by parts and taking into account the constant term $-1/4$ in eq.(\ref{log}), 
we finally obtain the result:
\be
\ln(R_N)= -\frac{1}{4}\ln(N)+\frac{1}{4}\int_0^{\infty}\frac{dt}{t}\left(
e^{-4t}-\frac{1}{(\ch(t))^2}\right)-\frac{1}{64~N^2}, 
\label{int}
\ee
where the omitted terms are of order $\sim 1/N^4$. One can see that the above 
expression coincides with the formula proposed in ref.\cite{Luk} in the case 
of the XX - chain. 
The last term in (\ref{int}) gives the following coefficient for the 
next-to-leading asymptotics for the correlator: 
\be
G(x)\simeq \frac{C_0}{\sqrt{\pi}}\left((-1)^x\frac{1}{x^{1/2}}
-\frac{1}{8}\frac{1}{x^{5/2}}\right),
\label{asympt}
\ee
where the constant $C_0$ is defined in (\ref{coas}). 
The value of the constant corresponing to the subleading term in eq.(\ref{asympt})
was first obtained by McCoy \cite{McCoy} using the method \cite{Wu} 
with the help of the asympotics of the Barnes $G$- function \cite{Barnes}, 
defined by $G(z+1)=\G(z)G(z)$, $G(1)=1$.  
The product $R_N$ can be represented as 
\[
  R_N=(G(1/2))^2\frac{(G(N+1))^2}{G(N+1/2)G(N+3/2)}.  
\]
Using the asymtotics of the function $G(N)$ at large $N$, 
\[   
     G(N)= \frac{1}{12}-\ln A - \frac{1}{2}(\ln2\pi)N + \left(\frac{1}{2}N^2
- \frac{1}{12}\right)\ln N - \frac{3}{4}N^2 +O\left(\frac{1}{N^2}\right), 
\]
where $A$ is the Glaishers constant (see below), one can obtain 
the result (\ref{asympt}) with the constant 
$(C_0)^{1/2}=\pi^{1/4}2^{1/12}e^{1/4}A^{-3}$. 
Note that using the asymptotics of $\psi(z)$ one can find analytically
the higher terms in the asymptotics of the correlator. 
One could also represent the first term in eq.(\ref{terms}) as 
\[
 -\frac{1}{2}\gamma+\sum_{p=2}^{\infty}
\frac{1}{p}\left(\frac{1}{2}\right)^{2p}\zeta(2p-1), 
\]
which also allows a numerical estimation of the constant. 
One can evaluate the integral in (\ref{int}) to get the asymptotic 
\[
\ln(R_N)= -\frac{1}{4}\ln(N)+ \left(\frac{\ln2}{12}+3\zeta'(-1)\right), 
\]
which is equivalent to the estimate 
\[
R_N=\left(N^{-1/4}\right)\left(2^{1/12}e^{1/4}A^{-3}\right), 
\]
where A- is the Glaisher constant 
\[
A=e^{1/12-\zeta'(-1)}=1.282427 \ldots. 
\]
This result agrees with the result obtained by Wu \cite{Wu}, using the 
expression of the product through the Barnes G- functions \cite{Barnes}.   
Finally, an alternative way to obtain the integral in eq.(\ref{int}) 
is to use the relations for $\G$ - function $\G(n+1)=n!$, 
$\G(n+1/2)=\pi^{1/2}2^{-n}(2n-1)!!$ to represent $R_N$ in the form:
\[
R_N= \prod_{k=1}^{N}\frac{\left(\G(k)\right)^2}{\G(k+1/2)\G(k-1/2)}. 
\]
Then one can use the known integral representation for the sum of the type 
$\sum_{k=1}^{N}\ln\G(k+b)$ (see.\cite{Brychkov}), and extract the 
large $N$ asymptotics of the resulting expression. 
Thus for the constant before the asymptotic (\ref{coas}) the value 
$C_0/2\sqrt{\pi}=0.147088\ldots$ is obtained. 

\vspace{0.2in}

{\bf Conclusion.}

\vspace{0.2in}

  In conclusion, we verified the functional form of the asymptotics of the spin-spin 
equal -time correlation function for the XX-chain predicted by different methods. 
We find the excellent agreement of the exact correlator with the prediction given 
by the leading asymptotics result up to the very small distances. We have also 
estimated the coefficient corresponding to the subleading correction \cite{McCoy}, 
\cite{Wu} in a different way and found the expression for the leading term in 
agreement with ref.\cite{Luk}. 
 It is worth mentioning that the correlators for the XY- spin chain 
and the Ising- like spin chain have been studied at the non-zero temperature and 
in the time - dependent case in different limits in a number of papers 
(for example, see ref.\cite{Tracy}).   
Let us mention that the exact correlator in the XX- model 
(at least in the infinite $L$ limit)  can be obtained in the framework of the 
general approach based on the Algebraic Bethe Ansatz method (see for example 
\cite{K}) as well as in the approach based on the formfactors of the model \cite{Colomo}
which is beyond the scope of the present letter. Let us mention also that 
studying of the formfactors with the small energy of the excited states can 
be the basis of the determination of the constant $C_0$. We postpone this question 
for a separate publication. It is possible that the formfactor approach 
can be usefull for the calculation of the constant $C_0$ in the general case of 
the XXZ - spin model.


\begin{thebibliography}{99}

\bibitem{Cardy}
J.Cardy, Nucl.Phys. B 270 (1986) 186. 
\bibitem{ML}
D.Mattis, E.Lieb, J.Math.Phys. 6 (1965) 304. 
\bibitem{M}
S.Mandelstam, Phys.Rev. D 11 (1975) 3026.\\
S.Coleman, Phys.Rev. D 11 (1975) 2088.
\bibitem{LP}
A.Luther, I.Peschel, Phys.Rev.B 9 (1974) 2911;~ Phys.Rev.B 12 (1975) 3908.
\bibitem{LSM}
E.Lieb, T.Schultz, D.Mattis, Ann.Phys. 16 (1961) 407.
\bibitem{McCoy}
B.M.McCoy, Phys.Rev. 173 (1968) 531;  \\
E.Barouch, B.M.McCoy, Phys.Rev. A 3 (1971) 786.
\bibitem{Luk}
S.Lukyanov, A.Zamolodchikov, Nucl.Phys. B 493 (1997) 571. 
\bibitem{Haldane}
F.D.M.Haldane, Phys.Rev.Lett. 47 (1981) 1840;~ J.Phys.C 14 (1981) 2585~;
Phys.Rev.Lett. 45 (1980) 1358.  
\bibitem{Wu}
T.T.Wu, Phys.Rev. 149 (1966) 380.  
\bibitem{Kar}
H.J.de Vega, M.Karowski, Nucl.Phys.B 285 (1987) 619;~
F.Woynarovich, H.P.Eckle, J.Phys.A 20 (1987) L97;~ 
M.Karowski, Nucl.Phys.B 300 (1988) 473;~ 
F.C.Alcaraz, M.N.Barber, T.M.Batchelor, Phys.Rev.Lett. 58 (1987) 771. 
\bibitem{Barnes}
E.W.Barnes, Quart.J.Math. 31 (1900) 264. 
\bibitem{Brychkov}
Yu.Prudnikov, A.Brychkov, O.Marichev, Integrals and Series, Moscow, Nauka, 1981. 
\bibitem{Tracy}
H.G.Vaidya, C.A.Tracy, Physica A 92 (1978) 1; 
U.Brandt, K.Jacoby, Z.Physik B 26 (1977) 245;
J.H.H.Perk, H.W.Capel, G.R.Quispel, F.W.Nijhoff, Physica A 123 (1984) 1;
J.H.H.Perk, H.W.Capel, Physica A 89 (1977) 265; 
T.T.Wu, B.M.McCoy, C.A.Tracy, E.Barouch, Phys.Rev.B 13 (1976) 316.  
\bibitem{K}
V.E.Korepin, Commun.Math.Phys. 94 (1984) 93; 
A.G.Izergin, V.E.Korepin, Commun.Math.Phys. 94 (1984) 67;  
F.H.L.Essler, H.Frahm, A.G.Izergin, V.E.Korepin, Commun.Math.Phys. 174 (1995) 
191.  
\bibitem{Colomo}
F.Colomo, A.G.Izergin, V.E.Korepin, V.Tognetti, Theor.Math.Fiz. 94 (1993) 19. 



\end{thebibliography}
\end{document}